\documentclass[aps,prl,twocolumn,superscriptaddress,showpacs]{revtex4}%
\usepackage{amssymb}
\usepackage[dvips]{graphics}
\usepackage{amsmath}
\usepackage{amsfonts}
\usepackage{graphicx}%
\usepackage{color}

\usepackage{graphicx}
\usepackage{dcolumn}
\usepackage{bm}

\begin{document}

\preprint{APS/123-QED}
\title{Intrinsic gap and exciton condensation in the $\nu_{\mathrm{T}}=1$ bilayer system}

\author{P. Giudici}
\altaffiliation[Present address: ]{Fakult\"at f\"ur Physik,\, Universit\"at Regensburg,\, 93040\, Germany.}
\affiliation{NTT\,Basic\,Research\,Laboratories,\,NTT\,Corporation,\,3-1\,Morinosato-Wakamiya,\,Atsugi\,243-0198,\,Japan}
\author{K. Muraki}
\affiliation{NTT\,Basic\,Research\,Laboratories,\,NTT\,Corporation,\,3-1\,Morinosato-Wakamiya,\,Atsugi\,243-0198,\,Japan}
\author{N. Kumada}
\affiliation{NTT\,Basic\,Research\,Laboratories,\,NTT\,Corporation,\,3-1\,Morinosato-Wakamiya,\,Atsugi\,243-0198,\,Japan}
\author{T. Fujisawa}
\altaffiliation[Present address: ]{Tokyo Institute of Technology, 2-12-1 Ookayama, Meguro, Tokyo 152-8551, Japan}
\affiliation{NTT\,Basic\,Research\,Laboratories,\,NTT\,Corporation,\,3-1\,Morinosato-Wakamiya,\,Atsugi\,243-0198,\,Japan}

\date{\today}

\begin{abstract}
We investigate the quasiparticle excitation of the bilayer quantum Hall (QH) system at total filling factor $\nu_{\mathrm{T}} = 1$ in the limit of negligible interlayer tunneling under tilted magnetic field. We show that the intrinsic quasiparticle excitation is of purely pseudospin origin and solely governed by the inter- and intra-layer electron interactions. A model based on exciton formation successfully explains the quantitative behavior of the quasiparticle excitation gap, demonstrating the existence of a link between the excitonic QH state and the composite fermion liquid. Our results provide a new insight into the nature of the phase transition between the two states.
\end{abstract}

\pacs{Valid PACS appear here}
\maketitle


The bilayer two-dimensional electron system at total filling $\nu_{\mathrm{T}} = 1$, with the lowest spin-split Landau level in each layer half filled, possesses an incompressible quantum Hall (QH) state in the regime of strong interlayer correlations~\cite{girvin}. 
This new state can be distinguished by a finite charge excitation gap from its compressible counterpart, a Fermi liquid state of composite fermions (CFs) with zero (or weak) interlayer correlations. This state is also unique in that, unlike other QH states, it possesses a broken symmetry in the absence of interlayer tunneling and can be viewed as a pseudospin ferromagnet~\cite{girvin} with the pseudospin encoding the layer degree of freedom or as an exciton condensate~\cite{kellogg2004,tutuc,wiersma} with excitons being formed from electrons and holes confined to different layers.

The capability of tuning the strength of the interlayer interactions by changing the electron density via gate voltages provides the unique opportunity to explore the $\nu_{\mathrm{T}} = 1$ system through its transformation between the weak and strong interaction limits. Experiments have indeed shown a phase transition between the compressible Fermi liquid and incompressible QH state as a function of $d/\ell_B$, the ratio between the inter- and intra-layer electron distance ~\cite{murphy,kellogg2004,wiersma}. This phase transition is usually interpreted in the mean-field theory as being driven by the charge density instability associated with the softening of a collective excitation mode~\cite{fertig}. However, experimental evidence of this mode softening is still lacking and the exact mechanism of the phase transition and the nature of the $\nu_{\mathrm{T}} = 1$ QH state at intermediate values of $d/\ell_B$ are still controversial~\cite{theo}. A recent experiment has shown that in the standard experimental conditions the compressible state is not fully spin polarized, which makes the phase boundary strongly dependent on the Zeeman energy~\cite{Bll}. This suggests that the behavior of the QH state in the vicinity of the phase transition reported thus far, including that in energy gap~\cite{wiersma}, tunneling~\cite{spielman2000,champagne2008}, drag~\cite{Kellogg2003}, light scattering~\cite{luin,karmakar2008}, and nuclear spin relaxation~\cite{spielman2005,kumada} experiments, is largely influenced by the partial spin polarization of the competing phase. Although the result of Ref.~\cite{Bll} has clarified how the phase boundary depends on the Zeeman energy, it does not provide any information about the properties of the $\nu_T = 1$ QH state, such as the nature of the quasiparticle excitation and how the system would behave in the absence of the spin degree of freedom. 

In this Letter, we investigate the physical origin of the quasiparticle excitation gap of the $\nu_{\mathrm{T}} = 1$ QH state and thereby clarify the intrinsic properties of the phase transition, through activation measurements in a tilted magnetic field. We demonstrate that, in the small tunneling regime, the quasiparticle excitation is of purely pseudospin character with no real spin involved.
When the spin degree of freedom is entirely suppressed at high fields, we obtain the intrinsic behavior of the gap, solely governed by $d/\ell_B$, toward the spin-independent phase transition. We develop a simple and quantitative model for the gap based on exciton condensation, which remarkably agrees with the measured gap without the need for adjustable parameters. Furthermore, the excitonic QH state is formed in our model from the composite fermion liquids, suggesting a close link between the competing states. Our results shed new light on the nature of the phase transition and QH state at intermediate $d/\ell_B$ for systems with and without spin degree of freedom.

The sample consists of two 18-nm-wide GaAs wells
separated by a 10-nm-thick AlAs/GaAs superlattice barrier. We take for the
interlayer distance $d$ the well center-to-center distance of
$28$~nm, which has an uncertainty of 5\%. By applying front- and back-gate biases, the electron density in each layer can be controlled. We work at the balance condition, with equal densities $n_{\mathrm{T}}/2$ in the two layers. The intralayer distance (or magnetic length) $\ell_B$ can be tuned using the relation $\ell_{B}=(2\pi n_{\mathrm{T}})^{-1/2}$ that holds at $\nu_{\mathrm{T}} = 1$. The structure has negligible tunneling, as confirmed by transport measurements through an individual layer, with the tunneling gap calculated to be $\Delta_{\mathrm{SAS}} = 150~ \mu$K~\cite{Bll}.
An in-plane field $B_{\parallel} = B\sin{\theta}$ was introduced by tilting the sample so that its normal forms an angle $\theta$ with the total field $B$. Measurements were performed at temperatures between $T=$~45 and 500 mK.


%
\begin{figure}[tb]
\resizebox{0.9\columnwidth}{!}{\includegraphics*{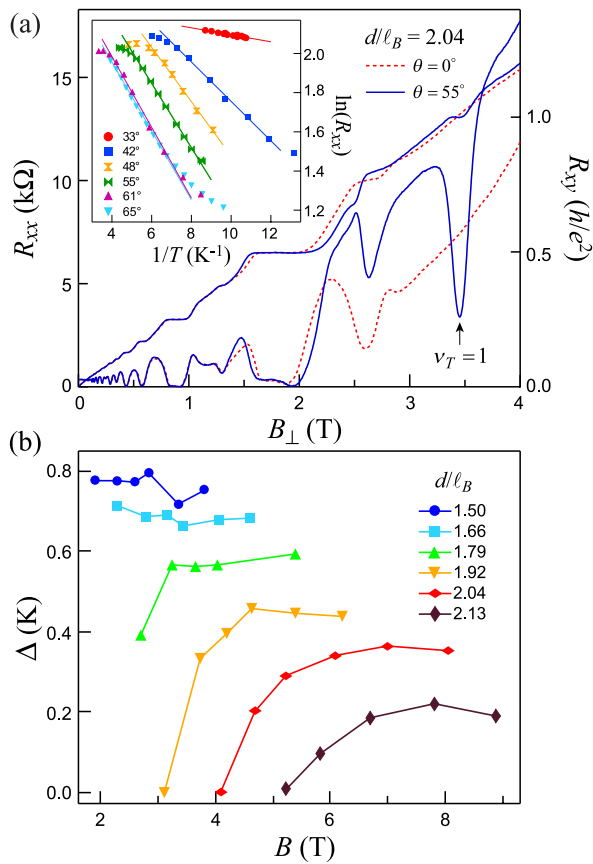}}
\caption{(color online). (a) Longitudinal $R_{xx}$ and Hall $R_{xy}$ resistances as a function of the perpendicular magnetic field $B_{\perp}$ for two tilt angles $\theta=0^{\circ}$ (dashed lines) and $55^{\circ}$ (solid lines), at the fixed $d/\ell_{B}= 2.04$. Inset: activation measurements of $R_{xx}$ at $d/\ell_{B}=2.04$  for six different tilt angles between $\theta=33^{\circ}$ and $65^{\circ}$.
(b) Quasiparticle excitation gap $\Delta$ as a function of total magnetic field $B$, for $\theta=0^{\circ}$, $33^{\circ}$, $42^{\circ}$, $48^{\circ}$, $55^{\circ}$,  $61^{\circ}$ and $65^{\circ}$. Different symbols correspond to different values of $d/\ell_{B}$. } \label{fig1}
\end{figure}

In bilayer systems with negligibly small $\Delta_{\mathrm{SAS}}$ the effect of $B_{\parallel}$ on the orbital part of the single-particle wave function is insignificant~\cite{Bll}, allowing the in-plane field to be used to change the Zeeman energy without modifying other parameters of the system. In the $\nu_{\mathrm{T}} =1$ system an increment of the Zeeman energy makes the compressible state, which is only partially polarized at low fields, energetically unfavorable, resulting in a shift of the phase boundary with the QH state to higher values of $d/\ell_{B}$~\cite{Bll}. 
This is illustrated in Fig.~1(a) by comparing the longitudinal ($R_{xx}$) and Hall ($R_{xy}$) resistances taken at $\theta=0^{\circ}$ and $55^{\circ}$ for a fixed $d/\ell_{B}=2.04$, plotted as a function of the perpendicular field $B_{\perp}$. While the $\nu_{\mathrm{T}}=1$ QH state is absent at $\theta=0^{\circ}$ due to the large value of $d/\ell_{B}$ (the transition occurs in this sample at $d/\ell_{B}=1.90$), the state is restored when tilting the sample at $\theta=55^{\circ}$.
As we show below, this enables us to investigate the intrinsic properties of the $\nu_T = 1$ QH state in a wide range of $d/\ell_{B}$.

\begin{figure}
\resizebox{0.9\columnwidth}{!}{\includegraphics*{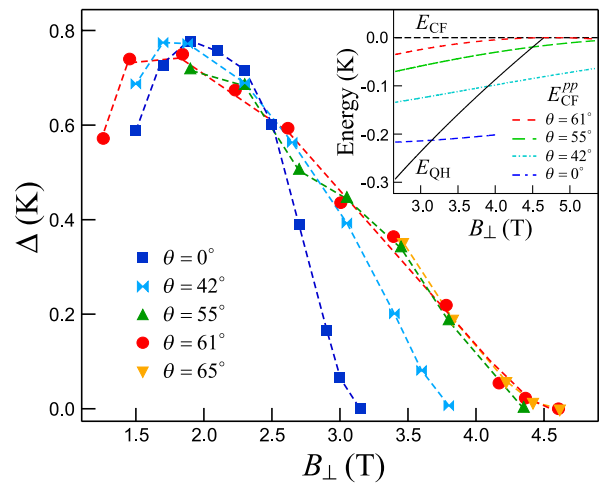}}
\caption{(color online).
Quasiparticle excitation gap $\Delta$ as a function of $B_{\perp}$, measured  at different tilted angles. 
Inset: calculated energies of the QH (solid line) and partially polarized compressible states (dashed lines), $E_{QH}$ and $E_{CF}^{pp}$ respectively, the latter for different angles. The energies refer to the fully polarized compressible state $E_{CF}$(horizontal line).} \label{fig3}
\end{figure}

The inset of Fig.~1(a) shows activation measurements of $R_{xx}$ at $d/\ell_{B}= 2.04$, taken at six different tilt angles between $\theta = 33^{\circ}$ and $65^{\circ}$. The quasiparticle gap $\Delta$, obtained by fitting the linear region using the relation $R_{xx}\varpropto\exp(-\Delta/2T)$, is plotted in Fig. 1(b) as a function of $B$ for different values of $d/\ell_{B}$. 
At $d/\ell_{B}=2.04$ a finite gap appears at $B\simeq 4.0$~T, which sharply rises with tilt, in accordance with the emergence of the QH state shown in Fig. 1(a). 
Similar behavior is observed for all data at $d/\ell_{B} \gtrsim$ $1.9$. The emergence of a finite gap is due to the shift of the phase boundary to higher $d/\ell_{B}$, which in turn brings the system into the QH phase. Indeed, the values of $d/\ell_{B}$ and $B$ at the onset of a finite gap are consistent with the previously reported phase diagram \cite{Bll}. Important here are that the strong change in the gap is confined to the vicinity of the phase transition and the size of the gap saturates at higher angles. Moreover, at low densities ($d/\ell_{B}\lesssim 1.7$), where the Zeeman energy is small, the gap is independent of the total field, clearly demonstrating that the quasiparticle excitation does not involve spin flips. This ensures that spin is irrelevant also at higher fields, where the energy cost due to spin flip becomes larger. We note that previous experiments have shown enhanced nuclear spin relaxation in the vicinity of $\nu_{\mathrm{T}}=1$  \cite{spielman2005,kumada}, which suggested a possible contribution of the real spin to the quasiparticle excitation~\cite{ps-spin}. With our results, however, in the small tunneling regime a Zeeman origin of the gap away from the phase transition can be definitely ruled out.

Having established the spin-independent origin of the gap, we now examine its dependence on $d/\ell_{B}$. In Fig.~\ref{fig3} we plot the measured gap against $B_{\perp}$ for $\theta=0^{\circ}$, $42^{\circ}$, $55^{\circ}$,  $61^{\circ}$ and $65^{\circ}$. The plot shows that for all angles the gap tends to zero upon increasing $B_{\perp}$. As the tilt angle is increased, the transition point gradually shifts to higher $B_{\perp}$ until $\theta = 61^{\circ}$, where the compressible state at the transition point becomes fully polarized and the phase boundary becomes Zeeman independent~\cite{Bll}.
Comparing the data at the high angles ($\theta=55^{\circ}$, $61^{\circ}$ and $65^{\circ}$) makes clear that the gap is angle independent and all data align on a single curve representing the intrinsic behavior of the gap, which depends only on $d/\ell_{B}$. Furthermore, the data at lower angles ($\theta=0^{\circ}$ and $42^{\circ}$) confirm that, except in the vicinity of the Zeeman-dependent phase transition, the gap follows the same curve. In what follows, we present a model for the spin-independent phase transition observed at high tilt and the associated gap formation.

The problem of the phase transition between the QH and compressible states is generally approached from the limit of strong interlayer interaction~\cite{fertig}; such an approach is useful in that one only needs to deal with the QH state whose properties are well known. Here we approach the problem from the opposite limit of weak interlayer interaction by assuming that the QH state is constructed from a system of independent layers through exciton formation. 
The inset of Fig. 3 illustrates the picture: as $d/\ell_B$ is decreased and the system enters the QH phase, a gap opens symmetrically around the energy of a state without interlayer correlations. Our assumption, which we test through comparison with experiment, is that this non-correlated state is the fully polarized compressible state and the QH state is formed from it through correlations. The energy gap is thus related to the energy gain due to exciton formation such that $\Delta=2E_{gain}$, where $E_{gain}=E_{CF}-E_{QH}$ is the difference between the compressible and QH state energies. We further assume that the non-interacting state, consisting of two nearly independent layers at $\nu=1/2$, can be effectively mapped onto a Fermi liquid of CFs at zero effective magnetic field~\cite{halperin}, of energy $E_{CF}=\pi\hbar^{2}n_{\mathrm{T}}/2m_{CF}$, with $m_{CF}$ the effective mass of CFs. On the other hand, the energy of the QH state is represented by the intra- ($\propto\ell_{B}^{-1}$) and inter- ($\propto d^{-1}$) layer Coulomb energies, and the gap becomes
\begin{equation}
\Delta =2(E_{CF}-\alpha\frac{e^{2}}{4\pi\epsilon\ell_{B}}-\beta\frac{e^{2}}{4\pi\epsilon
d}), \label{eq1}%
\end{equation}
where $\alpha$ and $\beta$ are some prefactors and $\epsilon$ is the dielectric constant of GaAs.
Taking the Coulomb energy $E_{C}=e^{2}/4\pi\epsilon\ell_{B}$ as a unit of energy, Eq.~(\ref{eq1}) can be written in the following dimensionless form:
\begin{equation}
\frac{\Delta}{E_C}=C-2\alpha-\frac{2\beta}{d/\ell_{B}},
\label{eq2}%
\end{equation}
where  $C=2E_{CF}/E_{C}$ becomes a constant because $m_{CF} \propto \sqrt{B_{\perp}}$~\cite{park}.
The values of $\alpha $, $\beta $, and $C$ (respectively, $0.024(16)\pm0.001$, $-0.025(63)\pm0.002$, $0.026(33)\pm0.001$) were separately determined in a previous work by fitting $d/\ell_{B}$ at the phase transition as a function of the normalized Zeeman energy~\cite{Bll}. Eq.~(2) is then a given function of $d/\ell_{B}$ and can be compared with the experimental data without adjustable parameters~\cite{parameters}.

\begin{figure}
\resizebox{0.8\columnwidth}{!}{\includegraphics*{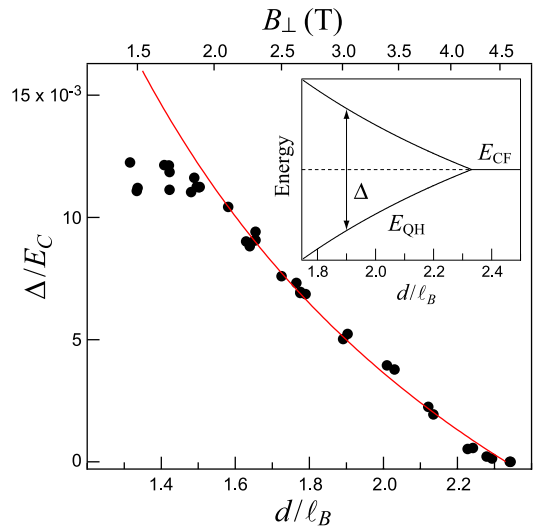}}
\caption{(color online).
Normalized gap $\Delta/E_C$ as a function of $d/\ell_{B}$ measured at $\theta=0^{\circ}$, $33^{\circ}$, $42^{\circ}$, $48^{\circ}$, $55^{\circ}$,  $61^{\circ}$ and $65^{\circ}$. The solid line is the gap obtained from the model (without fitting parameters). The inset shows how the gap $\Delta$ opens around the Fermi level ($\equiv E_{CF}$) with decreasing $d/\ell_{B}$.} \label{fig2}
\end{figure}

Figure~\ref{fig2} shows the comparison between the model and the experimental data, as a function of $d/\ell_{B}$. The experimental data are obtained from the saturated region of Fig.~1(b), at seven tilt angles between $\theta=0^{\circ}$ and $65^{\circ}$. 
The excellent agreement is more than intriguing: the precise relation between the gap and the energy gain implies that the QH state is constructed from the CF liquids by some mechanism, e.g., through exciton formation. Moreover, the continuous transformation between the two states implicit in the model evidences a second-order phase transition. The quantitative agreement obtained from our model is even more interesting if we consider that the charged excitation in the bilayer $\nu_{\mathrm{T}}=1$ condensate is believed to be a pseudospin texture that does not fit within a simple mean-field picture~\cite{girvin}. The remarkable agreement obtained from our results offers new elements for theoretical discussion. In the limit of $d/\ell_B \rightarrow 0$, the divergence of Eq.~(2) can be avoided by assuming the interlayer Coulomb energy to be proportional to $1/\sqrt{d^2+\ell_B^2}$ instead of $1/d$. In this case, equally good agreement is obtained over the same range of $d/\ell_B$, but with slightly different values of $\alpha$, $\beta$, and $C$. 
The disagreement between the model and the data at low values of $d/\ell_B$ is, however, not solved by this improvement. We ascribe it to disorder effects, which become important at low densities. 

Another important point is that in the above analysis the gap was calculated using the energy of the fully polarized CF liquid in the \emph{entire} range of $B_{\perp}$, whereas the energy of the partially polarized (pp) CF liquid, $E_{CF}^{pp}$, plays no role in the model. The inset of Fig.~2 compares  $E_{CF}^{pp}$ calculated for different $\theta$~\cite{note} with the energy of the fully polarized CF liquid $E_{CF}$ and the QH state. The diagram shows that $E_{CF}^{pp}$ strongly depends on $\theta$ and is lower than $E_{CF}$. Accordingly, the intrinsic spinless transition at $B_{\perp} \sim 4.5$~T is preempted by a transition to a pp-compressible state. Consistently with the level crossing shown in the inset, for $\theta<61^{\circ}$ the measured gap deviates from the intrinsic value, going to zero at the transition point to the pp-compressible state. Clearly, in this region no analogous relation as Eq.~(1) holds between the measured gap and $2(E_{CF}^{pp}-E_{QH})$, indicating that the pp-CF liquid plays no role in the formation of the gap and the QH state. This is important since it demonstrates that the transition to the pp-compressible state has a different nature than the intrinsic transition to the fully polarized compressible state: While the spinless transition is within the condensation picture (i.e., second order), a level crossing between states with different spin character (belonging to independent Hilbert subspaces) points to a first-order phase transition. Notably, this observation is also consistent with the results of recent experiments performed in the spin-dependent regime~\cite{karmakar2008}. 

Our results on the energy gap have revealed that the property of the $\nu_{\mathrm{T}}=1$ QH state measured in the vicinity of the spin-dependent transition does not reflect the intrinsic property of the condensate, being largely affected by the pp-compressible state that lies close in energy. 
According to the scenario of a first-order transition, one would expect the gap to drop discontinuously to zero at the phase boundary. Indeed, the gap decreases more rapidly with $B_{\perp}$ when the transition is spin dependent (e.g. $\theta = 0^{\circ}$), but not as abruptly as expected for a first-order transition. We ascribe this broadening to disorder, which leads to the phase coexistence of the QH and pp-compressible states over a finite range of $d/\ell_{B}$ (or $B_{\perp}$). 
This supports proposed models~\cite{fertig2005} for interpreting some behavior of the tunneling conductance~\cite{spielman2000} and counter flow conductivity \cite{kellogg2004} in terms of phase coexistence, probably induced by spatial variations in the interlayer distance and/or electron density. In such a situation, the presence (absence) of a finite gap in transport indicates the percolation of the QH (compressible) phase over the thermal length. The reduced gap in this "transition" region may be due to finite widths of the incompressible strips, which would reduce the pseudospin stiffness below its bulk value.
In turn, our results suggest that such phase coexistence is absent at high tilt angles where the energy gap takes its intrinsic value, which would allow the intrinsic property of the system to be investigated down to the quantum critical point of the continuous transition.

In summary, we demonstrated that the quasiparticle excitation is of purely pseudospin origin and associated with the gap that opens during exciton condensation. On one hand, the precise relation observed between the energy gap and the energy gain of the condensate with respect to the fully polarized compressible state attests to the underlying link between the two states, strongly suggesting a phase transition of second order. When the compressible state is only partially polarized, on the other hand, such a relation does not hold, demonstrating that the transition has a different character in this case, most likely first order. Our results definitely close the issue of the inconsistency between previous experiments in the spin-dependent regime and theories investigating an idealized spinless system.

\begin{acknowledgments}
The authors thank A. H. MacDonald for fruitful discussions.
\end{acknowledgments}


\begin{thebibliography}{99}
\bibitem{parameters} If we use $C-2\alpha$ and $\beta$ as adjustable parameters instead, and fit the data using Eq.~(2) in the range $d/\ell_B > 1.55$, we obtain $C-2\alpha = -0.022147 \pm 0.0002$ and $\beta = -0.025781 \pm 0.0002$, values that agree well with those adopted from Ref.~\cite{Bll} and used in Fig.~3 for the comparison with experiment.
\bibitem{note} More details about the model containing the spin degree of freedom can be found Ref.~\cite{Bll}.
\end{thebibliography}
\end{document}